\documentclass[twoside,twocolumn,english,aps,prb,showpacs,10pt]{revtex4-1}
\usepackage{babel}
\usepackage{graphics}
\usepackage{dcolumn}
\usepackage{amsmath}
\usepackage{stmaryrd}
\begin{document}
       
\title{Origin of the optical second-harmonic generation in spherical gold nanoparticles: local surface and non-local bulk contributions}

\author{G. Bachelier}
\email[Email: ]{guillaume.bachelier@lasim.univ-lyon1.fr}
\author{J. Butet, I. Russier-Antoine, C. Jonin, E. Benichou and P.-F. Brevet}
\affiliation{Laboratoire de Spectrom\'etrie Ionique et Mol\'eculaire (LASIM), Universit\'e Claude Bernard Lyon 1 - CNRS (UMR 5579), Bat. A. Kastler, 43 Bd du 11 novembre 1918, 69622 Villleurbanne, France.}

\begin{abstract}

The second-harmonic generation of 150 nm spherical gold nanoparticles is investigated both experimentally and theoretically. We demonstrate that the interference effects between dipolar and octupolar plasmons can be used as a fingerprint to discriminate the local surface and non-local bulk contributions to the second-harmonic generation. By fitting the experimental data with the electric fields computed with finite-element method (FEM) simulations, the Rudnick and Stern parameters weighting the relative nonlinear sources efficiencies are evaluated and the validity of the hydrodynamic model and the local density approximation approaches are discussed.

\end{abstract}


\maketitle

\section{Introduction}

Nanoscience and nanotechnology have focused a wide interest in the past decades on noble-metal nanostructures in view of the unique optical properties offered by the surface plasmon resonances (SPR). Their understanding is nowadays reaching one of its ultimate development with the ability to combine numerical simulations, 3D transmission electronic microscopy, optical extinction measurements or electron loss spectroscopy at the single particle level.\cite{Perassi2010,Nelayah2007,Muskens2008} In addition, new opportunities in plasmonics and meta materials have been recently opened with the coupling electric and magnetic dipoles in split-ring resonators or the tailoring of the SPR using Fano interferences.\cite{Shalaev2007,Bachelier2008Fano} Though, one of the main emerging challenges lies in the dynamical, time-resolved control of the SPR. In this context, one appealing route is the use of nonlinear optics driven by femtosecond laser pulses. The understanding of the nonlinear processes at play in plasmonic nanostructures is therefore becoming a very active topic.

The second-harmonic generation (SHG), whereby two photons at the fundamental frequency are converted into a single photon at the harmonic frequency, has quite a long history as it will soon celebrate its 50 years anniversary with the pioneer works of Franken et al.\cite{Franken1961} The origin of the nonlinear sources has early been established as arising from the breakdown of the centrosymmetry at the metal surface (local response) or from field gradients inside the bulk (non-local response). Two theoretical approaches were pursued to account for their relative intensity: the analytic hydrodynamic model introduced by J.E. Sipe et al.\cite{Sipe1980} and the density functional approach proposed by A. Liebsch.\cite{Liebsch1988} The later was found to be the most accurate description, although a complete theoretical framework involving both intraband and interband transitions is still missing. Noble-metal nanoparticles have triggered a renewed interest in SHG. Especially, the size dependence of the nonlinear efficiency, the effects of resonance, composition and chirality of the nanostructures have been addressed.\cite{Russier2007,Russier2008,Hao2002,Chandra2009,Kujala2007,Canfield2007,Valev2009} Single particle sensitivity has also been reached in different configurations.\cite{Butet2010,Jin2005,Canfield2008} Analytical models have been developed for simple geometries such as spheres and cylinders.\cite{Dadap1999,Beer2007,Valencia2004} If these important works mainly derived the selections rules and therefore the excited and radiating multipolar surface plasmon modes, they did not quantitatively evaluate the relative contributions of surface and bulk sources to the SHG, which therefore remains a largely open question. Few recent works involving numerical simulations have suggested that the bulk currents are indeed a fundamental contribution to the nonlinear response in noble-metal nanostructures.\cite{Zeng2009}

In this work, we investigate both experimentally and theoretically the nonlinear optical properties of 150~nm spherical gold nanoparticles in solution. We demonstrate for the first time that the interference effects between dipolar and octupolar plasmons, which has been addressed recently\cite{Butet2010sub}, can be used as a fingerprint to discriminate the local surface and non-local bulk contributions to the second-harmonic generation. In particular, the Rudnick and Stern parameters, weighting their relative efficiencies,\cite{Rudnick1971} are evaluated by fitting the experimental data with the electric fields computed with finite-element method (FEM) simulations.

\section{Results and discussion}
\subsection{Local surface and non-local bulk contributions}

For isotropic and centrosymmetric materials, the second-order surface susceptibility tensor can be drastically reduced to only three independent components: \(\chi_{\perp\perp\perp}\), \(\chi_{\perp\parallel\parallel}\) and \(\chi_{\parallel\parallel\perp}\) where \(\perp\) and \(\parallel\) stem for perpendicular and parallel to the surface, respectively. However, from both theoretical and experimental point of view,\cite{Wang2009} the \(\chi_{\perp\parallel\parallel}\) component only weakly contributes to the SHG of noble-metals. Hence, the two dominant surface nonlinear polarizations can be written as
\begin{equation}
\label{Pnnn}
P_{surf,\perp}(\mathbf{r},2\omega)=\chi_{\perp\perp\perp}E_{\perp}(\mathbf{r},\omega)E_{\perp}(\mathbf{r},\omega)
\end{equation}
\begin{equation}
\label{Pttn}
P_{surf,\parallel}(\mathbf{r},2\omega)=\chi_{\parallel\parallel\perp}E_{\parallel}(\mathbf{r},\omega)E_{\perp}(\mathbf{r},\omega)
\end{equation}
where the electric fields are evaluated inside the metal. These polarization vectors lead to surface currents located just outside the metal \cite{Sipe1980} and given by \(\mathbf{j}_{surf}=\partial\mathbf{P}_{surf}/\partial t\). From general considerations, the non-local bulk polarization may also have different contributions. However, in the special case of noble metals probed by a single fundamental field, it reduces to the following expression\cite{Wang2009}
\begin{equation}
\label{Pbulk}
\mathbf{P}_{bulk}(\mathbf{r},2\omega)=\gamma_{bulk}\mathbf{\nabla}.\left[\mathbf{E}(\mathbf{r},\omega).\mathbf{E}(\mathbf{r},\omega)\right],
\end{equation}
where \(\gamma_{bulk}\) is the bulk susceptibility. The associated bulk current is given by \(\mathbf{j}_{bulk}=\partial\mathbf{P}_{bulk}/\partial t\). Hence, the knowledge of \(\chi_{\perp\perp\perp}\), \(\chi_{\parallel\parallel\perp}\) and \(\gamma_{bulk}\) allows in principle to completely determine the nonlinear optical properties of noble-metals nanostructures. Following Sipe~\emph{et al.}\cite{Sipe1980} these parameters can be linked to the adimensional Rudnick and Stern parameters\cite{Rudnick1971} through the relations
\begin{equation}
\chi_{\perp\perp\perp}=-\frac{a}{4}\left[\epsilon_r(\omega)-1\right]\frac{e\epsilon_0}{m\omega^2}
\end{equation}
\begin{equation}
\chi_{\parallel\parallel\perp}=-\frac{b}{2}\left[\epsilon_r(\omega)-1\right]\frac{e\epsilon_0}{m\omega^2}
\end{equation}
\begin{equation}
\gamma_{bulk}=-\frac{d}{8}\left[\epsilon_r(\omega)-1\right]\frac{e\epsilon_0}{m\omega^2}
\end{equation}
where \(e\) and \(m\) are the electron charge and mass and \(\epsilon_0\epsilon_r(\omega)\) stems for the dielectric function of the metal at the frequency~\(\omega\).\cite{Johnson1972} In the framework of the hydrodynamic model,\cite{Sipe1980} the Rudnick and Stern parameters have very simple expressions: \(a=1\), \(b=-1\) and \(d=1\). However, if both tangential and bulk currents are well described by macroscopic parameters related to the dielectric functions, the normal current is driven by the local distribution of the electronic density over the Fermi length below the metal surface. Therefore, one has to go beyond the hydrodynamic model in order to properly evaluate the parameter \(a\). In particular, the latter was found to be larger than \(1\), frequency dependent and complex valued using local density approximation applied to conduction electrons only.\cite{Liebsch1988} It is also important to note that both the hydrodynamic model and the local density approximation approach were designed for planar surfaces and not for spherical particles. However, for particle radii by far larger than the Fermi length, the surface curvature effect can be assumed to be negligible, as well as quantum size effects.

\begin{figure}
\center
\resizebox*{7cm}{6cm}{\includegraphics{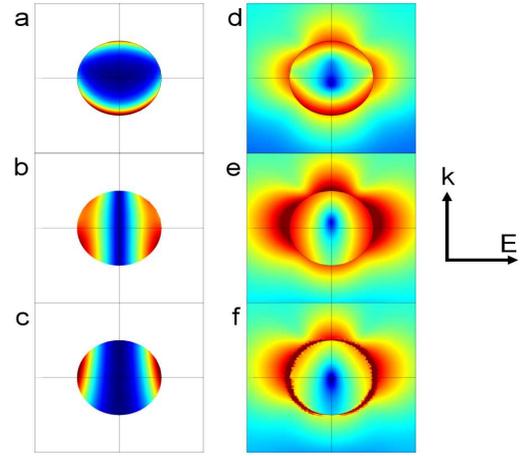}}
\caption{\label{Figure1}Spatial distribution of the nonlinear currents calculated for the bulk contribution (a), the tangential (b) and normal (c) surface currents for a 150 nm gold nanoparticle excited at 800nm. The near-field map of the corresponding harmonic field amplitudes are given in panels d-f.}
\end{figure}

\subsection{Finite-element method simulations}

The nonlinear currents corresponding to Eqs.~\ref{Pnnn}-\ref{Pbulk} were computed for a 150~nm gold nanoparticle assuming \(a=1\), \(b=-1\) and \(d=1\) (see Fig.~1a-1c). The fundamental electric field was obtained through finite-element simulations (FEM) using the scattered field formulation and perfectly matched layers in order to avoid any unwanted reflection at the simulation boundaries.\cite{Jin2002} In a second step, the near-field distribution of the harmonic field was calculated solving the Maxwell's equations within the weak formulation, allowing to incorporate both surface and bulk sources. As shown in Fig.~1d-1f, the obtained near-field maps around the nanoparticle do not strongly depend on the nonlinear source, except in terms of overall intensity. The latter is directly driven by the parameters \(a\), \(b\) and \(d\) but also by the evanescent and therefore localized character of the harmonic field. The knowledge of the far-field SHG intensity is a key ingredient to compare simulations with experimental measurements. It was computed from the near-field distribution of the harmonic field using the Stratton-Shu formula.\cite{Jin2002}

Figures 2a and 2b show the SHG intensities simulated for an excitation beam (wavevector \(\mathbf{k}\)) along the \(z\) axis and a collection at right angle (along the \(y\) axis). The fundamental electric field polarization angle with respect to the \(x\) axis is noted \(\gamma\). For \(\gamma = 0\) the polarization is along the \(x\) axis i.e. perpendicular to the scattering plane and \(\gamma = 90 \deg \) corresponds to a polarization along the \(y\) axis, i.e. in the scattering plane. Depending on the selected polarization for the harmonic field, different patterns are observed. For a detection polarization perpendicular to the scattering plane, a four lobe pattern is obtained whatever the origin of the nonlinear sources, arising either from the surface or from the bulk of the nanoparticle (all curves are superposed in Fig.~2a). Invoking spherical harmonics characterized by the angular momenta (\(l,m\)) with \(-l\leq m\leq l\), the latter property can be explained considering that i) the (\(\mathbf{k},\mathbf{E}(\mathbf{r},\omega))\) plane and the plane normal to it are symmetry planes for the harmonic field allowing only even values for \(m\), ii) the component of the field perpendicular to symmetry planes cancels out imposing \(m\neq0\) for detection polarization perpendicular to the scattering plane and iii) the fundamental field contains only \(m=\pm1\) terms leading to polarization vectors in Eqs.~\ref{Pnnn}-\ref{Pbulk} having \(|m|\leq 2\). Hence, a quadrupolar pattern associated to \(m=\pm 2\) is always observed, whatever the excited multipole given by \(l\). This detection configuration is therefore unlikely to discriminate the different contributions from the polarization measurements.

\begin{figure} 
\center
\resizebox*{7cm}{7cm}{\includegraphics{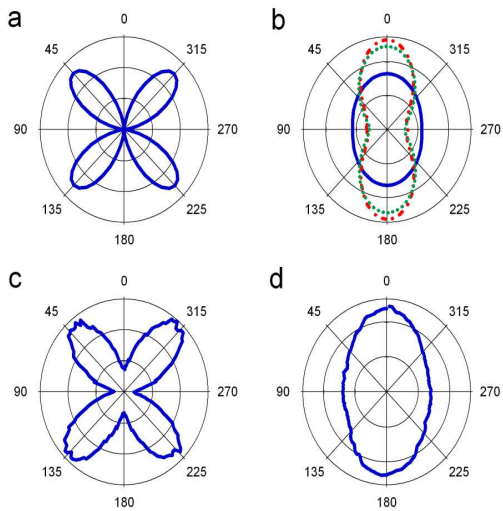}}
\caption{\label{Figure2}Normalized far-field second-harmonic intensity computed using FEM simulations (a and b) and measured experimentally (c and d) as a function of the electric field polarization at the fundamental frequency. The polarization of the harmonic field is perpendicular (a and c) and parallel (b and d) to the scattering plane. The normal surface current (dashed-dotted lines), tangential surface current (dotted lines) and bulk currents (full lines) contributions are shown separately.}
\end{figure}

\subsection{Discrimination between the nonlinear sources}

Owing to the surface and bulk origin of the nonlinear currents, it might be tentative to use the size dependence of the SHG intensity to assess the relative contributions of the nonlinear sources. As a matter of fact, it was found experimentally that the SHG intensity scales as the squared surface for small gold nanoparticles whereas a squared volume dependence is observed for larger ones.\cite{Russier2007} However, unexpected dipolar patterns, normally forbidden owing to the previous discussion, were revealed for small particles and for detection polarization perpendicular to the scattering plane. This was attributed to the centrosymmetry breaking of the particle shape\cite{Russier2007} as later demonstrated with FEM simulations.\cite{Bachelier2008} Indeed, all nonlinear sources lead to the same size dependence of the scattered intensity, regardless of their surface or bulk origin. This was verified with FEM simulations for spherical and deformed particles (data not shown here), but can be directly understood from Eqs.~\ref{Pnnn}-\ref{Pbulk}: the gradient appearing in the bulk polarization of centrosymmetric materials (Eq.~\ref{Pbulk}) introduces in spherical coordinates an additional term scaling as \(1/r\) and leading therefore to the same size dependence for the volume and surface polarization induced SH intensities. As a consequence, the size dependence cannot be used to separate bulk and surface contributions in centrosymmetric materials. 

Another strategy has therefore to be pursued in order to weight the nonlinear surface and bulk contributions in gold nanoparticles. This is provided by the interference effects between selected dipoles and octupoles we have recently demonstrated\cite{Butet2010sub} for a detection polarization in the scattering plane. In this specific configuration, constructive and destructive interferences are controlled by the incident polarization angle, leading to the varying intensities shown in Fig.~\ref{Figure2}b. The key point here is that the interference pattern depends on the nonlinear source (surface or bulk). More precisely, if both surface terms have nearly the same polarization dependence for the SHG intensity, they clearly deviate from the bulk contribution for which the intensity weakly oscillates for a 150~nm gold nanoparticle excited at 800~nm.  

\subsection{Quantitative determination of the nonlinear sources efficiencies}

The experimental data shown in Fig.~\ref{Figure2} were obtained using a modelocked Ti:sapphire laser tuned to a wavelength of 800 nm and delivering pulses of about 180 fs at a repetition rate of 76 MHz. The pulse energy measured at the laser exit was 10 nJ. The laser beam was focused onto a quartz cell with a microscope objective (x16, NA=0.32) leading to a beam waist at the fundamental frequency of \(5~\mu\)m.\cite{Butet2010} A low-pass filter was used in order to remove any residual light at the harmonic frequency generated prior to the cell. The SH photons are collected perpendicularly to the incident beam with a 25 mm focal length lens (NA=0.5). The scattered photons at the fundamental frequency are removed by a high-pass filter placed before the monochromator. The polarization angle of the fundamental beam is selected with a rotating half-wave plate and the polarization of the SH photons is selected by an analyzer. The photon detection was performed by a sensitive cooled photomultiplier tube and the fundamental beam was chopped at 130 Hz allowing a gated photon counting regime in order to remove the background light. The colloid solution of gold nanoparticles dispersed in water (1.4 pM) were purchased from BBI International (average diameter of 150~nm with a standard deviation of 8\(\%\)).

Comparing the calculated and measured signals (see Figs~2b and~2d), one can directly conclude that the normal surface current, which is usually considered as dominating the nonlinear response, cannot account by itself for the observed pattern, in contrast with the results obtained in Ref.~\onlinecite{Butet2010sub} for particle sizes smaller than 100~nm. Bulk and surface nonlinear sources being excited coherently, the corresponding SHG intensities shown in Fig.~2b cannot be summed up directly: the relative phases of the harmonic fields have to be taken into account. More precisely, it has to be noted that these phases depend on the input polarization angle since different multipoles are involved. This key feature allows the discrimination of the two surface contributions, having otherwise nearly the same intensity pattern (see Fig.~2b). The experimental data were therefore fitted by 
\begin{eqnarray}
I_{SHG}= && G \left|(a\mathbf{E}_{surf,\perp}(\mathbf{r},2\omega)+b\mathbf{E}_{surf,\parallel}(\mathbf{r},2\omega) \right. \nonumber \\ 
&&\left. +d\mathbf{E}_{bulk}(\mathbf{r},2\omega)\right|^2
\end{eqnarray}
where \(a\), \(b\) and \(d\) are the adjustable Rudnick and Stern parameters and \(G\) is a normalizing constant. \(\mathbf{E}_{surf,\perp}\), \(\mathbf{E}_{surf,\parallel}\) and \(\mathbf{E}_{bulk}\) are harmonic fields associated with the source terms given by Eqs.~\ref{Pnnn}-\ref{Pbulk} as obtained from the FEM simulations in the far-field region. The intensities recorded for a harmonic polarization in and out of the scattering plane were simultaneously fitted using the same Rudnick and Stern parameters in order to fully account for all experimental data. Owing to the normalizing constant \(G\), their absolute values cannot be extracted by the fitting procedure, but the relative amplitude and phase do. 

\begin{figure} 
\center 
\resizebox*{8.5cm}{12cm}{\includegraphics{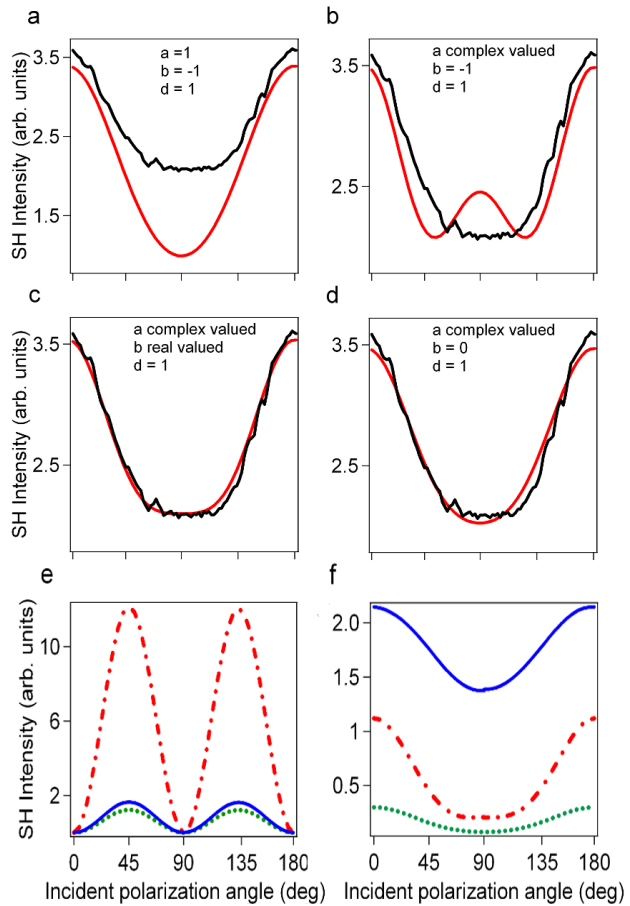}}
\caption{\label{Figure3}Fits of the experimental data (only the harmonic polarization in the scattering plane is shown) using the complex valued harmonic fields, as computed with the FEM simulations, and the following restrictions on the Rudnick and Stern parameters: (a) \(a=d=-b=1\), (b) \(a\) complex valued and \(d=-b=1\), (c) \(a\) complex valued, \(b\) real valued and \(d=1\) and (d) \(a\) complex valued, \(b=0\) and \(d=1\). The scattered intensities corresponding to the normal surface current (dashed-dotted lines), tangential surface current (dotted lines) and bulk currents (full lines) computed separately with the fitted Rudnick and Stern parameters of panel (c) are shown for a harmonic polarization parallel (e) and perpendicular (f) to the scattering plane.}
\end{figure}

\subsection{Comparison with existing models}

Different constraints were applied to the fitting Rudnick and Stern parameters in order to check the validity of existing models. We first considered the hydrodynamic model\cite{Sipe1980} where the Rudnick and Stern parameters satisfy \(a=d=-b=1\). Clearly, the obtained result cannot account for the experimental data, as shown in Fig.~3a. More advanced models were early proposed including a damping term to the hydrodynamic model\cite{Corvi1986} or taking into account the spatial distribution of the electrons over the Fermi length below the surface within local density approximation.\cite{Liebsch1988} All these models suggest that \(d=-b=1\) and \(a\) is complex valued and exhibit resonant behavior largely exceeding in amplitude the two other parameters. As a matter of fact, it was here again not possible to reproduce the experimental data by the fitting procedure (see Fig.~3b): additional minor peaks appear at 90~and 270~\(\deg\), in contrast with the experimental data. The fact that the damped hydrodynamic model and the local density approximation approaches can't accurately account for the SHG in gold nanoparticles is not surprising since they do not take into account the interband transitions that are resonantly excited at the harmonic wavelength (400~nm). Hence, a complete theoretical investigation of both intraband and interband transitions is required in order to correctly describe the nonlinear optical properties of noble-metal nanoparticles.

In order to go further in the understanding of the bulk and surface contributions in the SHG from gold nanoparticles, the condition \(d=-b=1\) was relaxed, allowing the parameter \(b\) to take any real values. As shown in Fig.~3c, a very good agreement between experimental data and FEM simulations is obtained for \(d=1\), \(b=0.1(1)\) and \(a=0.5(6)-i0.2(5)\). If either the volume contribution (\(d\)) or the tangential surface contribution (\(b\)) is suppressed, a less satisfactory fit is achieved as shown in Fig.~3d for \(b=0\): it is not possible to reproduce the plateau observed at 90~\(\deg\) nor the intensity maximums reached at 0~and 180~\(\deg\), indicating that some physics is missing. Hence all contributions are necessary to account for the scattered SH intensity, even if in the present case the tangential surface contribution is rather weak (see Fig. 3f). Despite the fact that \(|a|<|d|\), the normal surface current largely dominates the nonlinear response for a detection polarization perpendicular to the scattering plane (Fig.~3e), with an intensity nearly 7.5 larger than the other contributions.  In contrast, the SHG intensity collected for harmonic polarization in the scattering plane is dominated by the bulk currents as shown in Fig.~3f. The ratio between bulk and surface contributions ranges from 2 to 6.5 depending on the fundamental polarization angle. This clearly underlines the importance of the bulk source that can be the dominant contribution in the SHG from gold nanostructures depending on the experimental configuration.\cite{Zeng2009}

\section{Conclusion}

In conclusion, it is shown that the local surface and non-local bulk contributions to the harmonic generation can be discriminated using an interference effect between the selected dipolar and octupolar plasmon modes, in a specific scattering and polarization configuration. By fitting the experimental data by the simulated electric fields, we report the first quantitative determination of the nonlinear sources efficiencies in spherical gold nanoparticles. We show that the actual theoretical models, namely the hydrodynamic model and the density functional approach, cannot account for the experimental data obtained on 150~nm spherical gold nanoparticles. This is attributed to the interband transitions that are excited resonantly in the present case and not included in these models. Finally, we demonstrate that the relative local surface and non-local bulk contributions strongly depend on the experimental configuration.

\end{document}